\documentclass[a4paper,12pt]{article}
\usepackage{graphics} 

\date{}

\begin{document}

\title{ Minimal left-right symmetric models and new $Z'$ properties at future electron-positron colliders}
\author{ F. M. L. Almeida Jr., Y. A.
Coutinho, J. A. Martins Sim\~oes,\\ J. Ponciano, A. J. Ramalho and Stenio Wulck \\ Instituto
de F\'{\i}sica,\\
Universidade Federal do Rio de Janeiro, RJ, Brazil\\
and\\
M. A. B. Vale\\
Universidade Federal  de S\~ao Jo\~ao del Rei, MG, Brazil}
\maketitle
\begin{abstract}
\par
It was recently shown that left-right symmetric models for elementary
particles can be built with only two Higgs doublets. The general
consequence of these models is that the left and right fermionic sectors
can be connected by a new neutral gauge boson $Z'$ having its mass as the
only additional new parameter. In this paper we study the influence  of
the fundamental fermionic representation for this new neutral gauge boson.
Signals of possible new heavy neutral gauge bosons are investigated for
the future electron-positron colliders at $\sqrt s = 500$ GeV , $1$ TeV
and $3$ TeV.  The total cross sections, forward-backward and left-right
asymmetries and model differences are calculated for the process 
$e^+ e^- \longrightarrow \mu^+ \mu^-$. Bounds on $Z'$ masses are estimated.

\end{abstract}
\eject

\section{Introduction}\setcounter{equation}{0}

One possible way to understand the left-right asymmetry of elementary
particles is to enlarge the standard model into a left-right symmetric
structure and then, by some spontaneously broken mechanism, to recover the
low energy asymmetric world. There are three main points in this proposal:
the choice of the gauge group, the Higgs sector and the fundamental
fermionic representation.

\par
Left-right models starting from the gauge group $SU(2)_{L}\otimes SU(2)_{R}\otimes U(1)_{B-L}$ were developed by many authors \cite{JCP} and are well known to be consistent with the standard  $SU(2)_{L}\otimes U(1)_Y$ model.
This group can be part of more general models, like some grand unified groups \cite{BBR}, superstring inspired models \cite{ZGB}, a connection between parity and the strong $CP$ problem \cite{SMB}, left-right extended standard models \cite{RFO}. All these approaches imply the existence of some new intermediate physical mass scale, well bellow the unification or the Planck mass scale.
\par
For the Higgs sector there are some options. Two Higgs doublets that transform as fields in the left and right sectors can be supposed to be spontaneously broken at scales 
$v_{L} =v_{{Fermi}}$ and at a larger scale $v_R$ respectively. The earlier left-right symmetric models  added a new Higgs in the mixed representation $(1/2,\, 1/2,\, 0)$, for $(T_{3L},\,T_{3R},\,Y)$. The symmetry breaking of this field gives a mixing in the charged vector boson sector (not yet experimentally verified) and could also be responsible for neutrino masses. The increasing experimental evidence on neutrino oscillations and nonzero masses has  motivated a renewed interest in the mechanisms for parity breaking. More recently it was shown that all fermion masses could be obtained with only two Higgs doublets \cite{EMA}. The basic mechanism for this model is the dimension-5 operator built by Weinberg years ago \cite{WEI}. It is also possible to build mirror models with two Higgs doublets and new Higgs singlets \cite{NOS}. In this case charged fermion masses can be understood as a result of a see-saw mechanism. 
\par
Throughout this paper we call models with two Higgs doublets "minimal models" in the sense that they have the minimal set of new scale parameters that are shown to be consistent with the standard $SU(2)_{L}\otimes U(1)_{Y}$ theory.
\par
For the fermion  spectrum there is no unique choice of the fundamental fermionic representation. Earlier left-right models
restored parity by choosing the right-handed sector as doublets under $SU(2)_R$  with $\nu _R$ and $u_R$ as the upper components of the right doublet. Other models have doubled the number of fundamental fermions choosing the new sector with opposite chirality relative to the standard model sector.
\par
In this paper we present  models that start by the simple gauge structure of $SU(2)_{L}\otimes SU(2)_{R}\otimes U(1)_{B-L}$ and investigate the consequences of the minimal Higgs sector that breaks the left-right symmetry. This paper is organized as follows: in section 2 we review the main assumptions for the Higgs and gauge sector in the minimal left-right model; in section 3 we review the properties of new fermion representation; in section 4 we show some phenomenological consequences for testing the models here proposed and in section 5 we give our conclusions.

\section{The Higgs and gauge boson sectors in the minimal model}\setcounter{equation}{0}

\par
Left-right models with only two Higgs doublets have been previously considered \cite {EMA,NOS}. We review in this section the main points that are relevant for the new neutral current interactions. The minimal left-right symmetric model contains the following Higgs scalars:

\begin{eqnarray}
\begin{array}{ccc}
& \chi_L=\left(\begin{array}{cc}
\chi_L^+\\
\chi_L^0\end{array}\right),&
\chi_R=\left(\begin{array}{cc}
\chi_R^+\\
\chi_R^0\end{array}\right),
\end{array}
\end{eqnarray}

\noindent with transformation properties under $SU(2)_L\times SU(2)_R \times U(1)_Y$

\begin{equation}
 (1/2,0,1)_{\chi_L},\; (0,1/2,1)_{\chi_R}.
\end{equation}
The first stage of symmetry breaking occurs when $\chi_R$ acquires its
vacuum expectation value $<\chi_R>$, leaving a remnant $U(1)_{Y'}$ symmetry
coming from the $SU(2)_R\times U(1)_Y$ sector, whose generator is  given by the relation
$\frac{1}{2} Y'= T_{3R} + \frac{1}{2}Y $, with $Y=B-L$.
The breakdown to $U(1)_{em}$ is realized, at the scale 
$ v_L\simeq v_{Fermi}$, through the following vacuum expectation value:

\begin{eqnarray}
\begin{array}{ccc}
& <\chi_L>=\left(\begin{array}{cc}
0\\
v_L\end{array}\right).
\end{array}
\end{eqnarray} 
\par
In order to analyze the couplings of the additional neutral gauge
boson we rewrite here the free Lagrangian for the gauge fields and the  
piece containing the covariant derivatives of the scalar fields

\begin{eqnarray}
\mathcal{L} &=&\mathcal{L}_0+\mathcal{L}_D\\ 
\mathcal{L}_0 &=&-\frac{1}{4} F^{\mu\nu}F_{\mu\nu}-\frac{1}{2}
Tr[G_{L\mu\nu}^aG_L^{a,\mu\nu}]-\frac{1}{2}Tr[G_{R\mu\nu}^aG_R^{a,\mu\nu}]\\ 
\mathcal{L}_D &=&
(D_{\mu}\chi_L)^{\dagger}D^{\mu}\chi_L+(D_{\mu}\chi_R)^{\dagger}D^{\mu}\chi_R,
\end{eqnarray}
where
\begin{eqnarray}
F^{\mu\nu}&=&\partial^{\mu}B^{\nu}-\partial^{\nu}B^{\mu}\\
G^{a,\mu\nu}_{L,R}&=&\partial^{\mu}W^{a,\nu}_{L,R}-\partial^{\nu}W^{a,\mu}_{L,R}
+ ig_{L,R}[W^{a,\mu}_{L,R},W^{a,\nu}_{L,R}],
\end{eqnarray}
and 
\begin{eqnarray}
D_{\mu}\chi_{L,R}&=&\partial_{\mu}\chi_{L,R}+ig_{L,R}W_{L,R}\chi_{L,R}.
\end{eqnarray}
The gauge coupling constants related to the gauge group $SU(2)_L\otimes SU(2)_R\otimes U(1)_{B-L}$, are respectively $g_L$, $g_R$ and $g$.
When substituting the vacuum expectation values for the scalar fields
in $\mathcal{L}_D$, one obtains the gauge bosons mass
terms. Explicitly, the mass matrix for the neutral sector in the basis
$(W_L,\,W_R,\,B)$ is   
\begin{eqnarray}
M=\frac{1}{4}\left(\begin{array}{ccc}
g_L^2 v_L^2 & 0 &-g g_L v_L^2 \\
0 & g_R^2 v_R^2 & -g g_Rv_R^2 \\
-g g_Lv_L^2 & -g g_Rv_R^2 & g^2(v_L^2+v_R^2)
\end{array}\right)
\end{eqnarray}
The mass matrix $M$ is diagonalized by an orthogonal transformation 
$R$ which connects the weak fields ($W_L^{\mu},\,W_R^{\mu},\,B^{\mu}$) to the
physical ones ($Z^{\mu},\,Z'^{\mu},\,A^{\mu}$).
By direct calculation from the neutral mass matrix we can obtain an
analytic expression for $R$ in powers of $w=v_L/v_R$,
\begin{equation}\label{explicitR}
R=\left(\begin{array}{ccc} 
\cos \theta_W & w^2 \cos \beta \sin^2 \beta  & \sin \theta_W \\
-\sin \theta_W \sin \beta -\displaystyle{ \frac{w^2\cos^2 \beta \sin^3 \beta }{\sin \theta_W}}   & \cos \beta - w^2 \cos \beta \sin^4 \beta & \sin \beta \cos \theta_W \\
-\sin \theta_W \cos \beta + \displaystyle{ \frac{w^2\cos \beta \sin^4 \beta }{\sin \theta_W}} & -\sin \beta -w^2 \cos^2 \beta \sin^3 \beta  & \cos \beta \cos \theta_W\\
\end{array}\right)
\end{equation}

\par
In Eq.(\ref{explicitR}), the following relations were employed
\begin{equation}
\sin^2\theta_W=\frac{g^2_Rg^2}{\Lambda},
\;\; \sin^2\beta=\frac{g^2}{g_R^2+g^2},\;\; \sin\alpha= 
\frac{g^2\Lambda^{1/2}}{(g_R^2+g^2)^2}w^2 + O(w^3)
\end{equation}
with $\Lambda=(g_L^2g_R^2+g^2 g_L^2+g^2 g_R^2)$.
\bigskip
\par
In the limit $w=0$, the left and right sectors are decoupled and one
recovers the standard model gauge boson couplings.
The triple and quartic self-interactions terms contained in the kinetic terms $Tr[G^{\mu\nu}G_{\mu\nu}]_{L,R}$ are explicitly
\begin{eqnarray}
Tr[G^{a,\mu\nu}G^a_{\mu\nu}]& = &\partial^{\mu}W^{a,\nu}\partial_{\mu}W_{\nu}^a-\partial^{\nu}W^{a,\mu}
\partial_{\mu}W_{\nu}^a+2ig_{L,R}f^{abc}\partial^{\mu} 
W^{a,\nu}W^b_{\mu}W^c_{\nu}+  \nonumber \\
&+& \frac{g^2}{2}f^{abc}f^{alm}W^b_{\mu}W^c_{\nu}W^{l,\mu}W^{m,\nu}
\end{eqnarray}
where $f^{abc}$ is the totally antisymmetric tensor.
Using the mixing matrix $R$ and  taking the  physical  
charged fields as being

\begin{equation}
W^{\pm \mu}_{L,R}=\frac{1}{\sqrt{2}}(W_{L,R}^{1 \mu}\pm W_{L,R}^{2\mu}),
\end{equation}

\noindent the Feynman rules for the $W^{+}_{L,R}W^{-}_{L,R}X_j$ triple vertices are readily found: 
\begin{eqnarray}
\Gamma_{\lambda_1\lambda_2\lambda_3}^{abc}(k_1,k_2,k_3)&=&g_iR_{ij}f^{abc} 
[(k_1-k_2)_{\lambda_3}g_{\lambda_1\lambda_2}+ 
(k_2-k_3)_{\lambda_1}g_{\lambda_2\lambda_3} + \nonumber \\
&+& (k_3-k_1)_{\lambda_2}g_{\lambda_3\lambda_1}],
\end{eqnarray}
with $i=1, \, 2$, $g_1\equiv g_L$, $g_2\equiv g_R$ and the sub-index $j$
takes the values $1, \, 2$ or $3$ whenever $X_j$ is identified as 
$Z$, \, $Z'$ or $\gamma$ respectively.
In Table (1) we summarize the results for the couplings
factors $g_iR_{ij}$ using the standard  
parametrization of $R$ in terms of  $\sin \theta_W(s_{\theta W})$, $\sin\beta(s_{\beta})$
and $sin\alpha (s_{\alpha})$, which correspond to the mixing angles between $Z-A$, $Z'-A$
and $Z'-Z$ respectively.

\begin{table}[!t]\label{triplo}
\begin{center}
\begin{tabular}{|c|c|c|c|}
\hline \hline
\multicolumn{4}{|c|}{Couplings}  \\ \hline \hline
& $Z$ & $Z'$ & $\gamma$ \\ \hline
$W^{+}_LW^{-}_L$ & $g_Lc_{\theta W}c_{\alpha}$ & $g_Lc_{\theta_W}s_{\alpha}$
& $g_Ls_{\theta_W}$\\ \hline
$W^{+}_RW^{-}_R$ & $-g_R(s_{\alpha}c_{\beta}-c_{\alpha}s_{\theta
W}s_{\beta})$
 & $g_R (c_{\beta}c_{\alpha}-s_{\alpha}s_{\theta_W}s_{\beta})$ & $g_R
s_{\beta}c_{\theta_W}$\\  \hline
\end{tabular}
\end{center}
\caption{Triple couplings.}
\end{table}

\par

Similarly, for the quartic self-interaction term a straightforward calculation
for the $W^{+}_iW^{-}_iX_jX_k$ vertex yields the Feynman rules:

\begin{eqnarray}
\Gamma^{abcd}_{\lambda_1\lambda_2\lambda_3\lambda_4}&=&g_i^2R_{ij}R_{ik}
[f^{abe}f^{cde}(g_{\lambda_1\lambda_3}
g_{\lambda_2\lambda_4}-g_{\lambda_2\lambda_3} g_{\lambda_1}g_{\lambda_4})+
\\ \nonumber 
&+& f^{ace}f^{bde}(g_{\lambda_1\lambda_2}g_{\lambda_3\lambda_4}-g_{\lambda_3\lambda_2}g_{\lambda_1\lambda_4})
+f^{ade}f^{cbe}(g_{\lambda_1\lambda_3}g_{\lambda_2\lambda_4}-g_{\lambda_4\lambda_3}g_{\lambda1\lambda_2}).
\end{eqnarray}
The resulting couplings are summarized in Table (2).

\begin{table}[!t]\label{quarticvtx}
\begin{center}
\begin{tabular}{|c|c|c|c|}
\hline\hline
\multicolumn{4}{|c|}{Couplings}  \\ \hline \hline
& $\gamma Z'$ & Z'Z & Z'Z'\\ \hline
$W_L^{+}W_L^{-}$ & $g_L^2s_{\theta_W}c_{\theta_W}s_{\alpha}$ & 
$g_L^2c^2_{\theta_W}s_{\alpha}c_{\alpha}$
& $g_L^2c^2_{\theta_W}s^2_{\alpha}$\\ \hline
$W_R^{+}W_R^{-}$ & $g_R^2s_\beta c_{\theta_W}$ &
$g_R^2(s_{\alpha}c_{\beta} + c_{\beta}s_{\theta_W}s_{\beta})(s_{\alpha}s_{\theta_W}s_{\beta}-c_{\beta}c_{\alpha})$ &
$g_R^2(c_{\beta}c_{\alpha}-s_{\alpha}s_{\theta_W}s_{\beta})^2$\\ 
\hline
\end{tabular}
\end{center}
\caption{Quartic couplings.}
\end{table}
In the  high energy limit where  the symmetry breaking scales $v_R$ and $v_L$ can be neglected,  the theory is invariant under the parity operation $\cal P$, and we must have $g_L = g_R $. At lower energies  the running couplings  lead to different values of $g_L$ and  $ g_R $. However, in the region of the $Z'$ that we are considering this is a small effect and we will consider $g_L=g_R$. This simplification reduces the number of the arbitrary gauge coupling to two.
\par
One of the most interesting consequences of the minimal left-right symmetric model is that there is only one new scale parameter in the model, $v_R$, besides the usual standard model inputs. 
\smallskip

\vskip 1cm
\section{Models for the fermion representation}\setcounter{equation}{0}
\par
We present in this paper two possibilities for the fundamental fermionic representation.

{\subsection {Mirror left-right model}}

\par 
In this model \cite{NOS} (from now on called MLRM) we have new heavy fermions with opposite chirality relative to the present known fermions. The parity operation transforms the $SU(2)_{L}\buildrel {\rm P} \over \longleftrightarrow  SU(2)_{R}$ sectors, including the vector gauge bosons. For the other leptonic and quark families a similar structure is proposed. The charge generator is given by $ Q = T_{3L}+T_{3R}+Y/2 $.

\par
The fundamental representation for leptons in this model is: 

\begin{equation} 
{\ell}_L= {{\nu \choose e}_L, \qquad \nu_R, \qquad e_R,\qquad  L_R={N \choose E}_R},\qquad N_L,\qquad E_L
\end{equation} 

\par
For quarks we have,

\begin{equation} 
{u}_L= {{u \choose d}_L, \qquad u_R, \qquad d_R,\qquad  U_R={U \choose D}_R},\qquad U_L,\qquad D_L.
\end{equation} 

\bigskip

\begin{table}[!t]\label{states}
\begin{center}
\begin{tabular}{||c||c||c||c||c||}
\hline States &  $T_{3L}$ & $T_{3R}$ & $Y/2$  & $Q$ \\ 
\hline $\nu_L$ &  $1/2$ & $0$ & $-1/2$ & $0$ \\ 
\hline  $e_L$  & $-1/2$ & $0$ & $-1/2$ & $-1$ \\ 
\hline $N_R$ & $0$  & $1/2$ & $-1/2$ & $0$ \\ 
\hline  $E_R$  & $0$  & $-1/2$ & $-1/2$ & $-1$ \\ 
\hline  $u_L$ &  $1/2$ & $0$ & $1/6$ & $2/3$ \\ 
\hline  $d_L$ &  $-1/2$ & $0$  & $1/6$ & $-1/3$ \\
\hline  $U_R$ &  $0$ & $1/2$ & $1/6$ & $2/3$ \\ 
\hline  $D_R$ &  $0$ & $-1/2$  & $1/6$ & $-1/3$ \\ 
\hline 
\end{tabular} 
\end{center}
\caption {Quantum numbers for left and right states in mirror left-right model.}
\end{table}

\bigskip
\par 
The quantum numbers for this model are shown in Table (3) with the charge operator given by $Q= I_{3L}+I_{3R}+ \displaystyle{{B-L} \over 2}$ 
\par
Introducing the notation

\begin{eqnarray}
sin^2\theta_W  & \equiv & {{g^2_R g'^2}\over {g^2_R g^2_L +g^2_R g'^2 +g^2_L g'^2}}\nonumber\\
sin^2\beta & \equiv & {{g'^2}\over{g^2_R + g'^2}},
\end{eqnarray}

\noindent the condition $g_L=g_R$ implies 
\begin{equation}
 \sin\beta=\tan\theta_W 
\end{equation}
and the unification condition for the electromagnetic interaction is the same as in the standard model,

\begin{equation}
e= g_L \sin \theta_W.
\end{equation}
\par

We are interested in interactions between the extra neutral gauge boson $Z'$ and the ordinary
fermions, that are described by the Lagrangian for the neutral currents with 
$Z$ and $Z'$ boson contributions,

\begin{eqnarray}
{\cal L_{NC}}  = & & \frac{e}{4 \sin\theta_W \cos\theta_W}\bar \Psi_i \gamma^{\mu}\lbrace 
T_{3L}\frac{(1- \gamma_5)}{2}- Q \sin ^2 \theta_W \rbrace\Psi_i Z_{\mu} \nonumber\\ 
\nonumber\\
+ & & \frac{e \tan\theta_W \tan \beta}{4 \sin\theta_W} 
\bar \Psi_i \gamma^{\mu} \lbrace T_{3L}\frac{(1- \gamma_5)}{2}- Q \rbrace \Psi_i Z'_{\mu}. 
\end{eqnarray}

\bigskip
\par
The couplings between the gauge neutral bosons and the matter fields are explicitly shown in Table (4).
\par

In this model the charged fermion masses can also be understood as having
its origin in a see-saw mechanism. This new result comes from the choice of the fundamental fermionic representation and from new Higgs singlets that do not contribute to the gauge boson masses \cite{NOS}.

\par

\begin{table}[!t]\label{jujuba}
\begin{center}
\begin{tabular}{||c||c||c||}
\hline  Couplings &  $g_V$ & $g_A$ \\ 
\hline  $Z\nu\nu$ & $1$ & $1$ \\ 
\hline  $Zee$ & $-1+4 sin\theta_W$ & $1$ \\ 
\hline  $Zuu$ & $3-8 sin\theta_W$  & $-3$ \\ 
\hline  $Zdd$ & $-3+4sin\theta_W$ & $3$  \\ \hline 
\hline  Couplings &  $g'_V$ & $g'_A$ \\
\hline  $Z'\nu\nu$ & $1$ & $(cos^2 \theta_W-sin^2\theta_W)$  \\ 
\hline  $Z'ee$ & $-1+sin^2\theta_W$ &  $(cos^2\theta_W-sin^2\theta_W)$\\ 
\hline  $Z'uu$ & $3-8sin^2\theta_W$ & $+3(cos^2\theta_W-sin^2\theta_W)$ \\ 
\hline  $Z'dd$ & $-3+4sin^2\theta_W$ & $-3(cos^2\theta_W-sin^2\theta_W)$ \\ 
\hline 
\end{tabular} 
\end{center}
\caption{Couplings between the neutral gauge bosons $Z$ and $Z'$ and the ordinary fermions in mirror left-right model (first family).}
\end{table}
\newpage

{\subsection {Symmetric left-right model}} 
\par 
\smallskip

In this model (from now on called SLRM) a new right handed fermionic sector
appears as a doublet under the $SU(2)_R$ transformation \cite{JCP}.

\par
The fundamental representation for leptons and quarks of the gauge group $SU(2)_L\otimes SU(2)_R\otimes U(1)_{B-L}$ is: 

\begin{equation} 
{\Psi_L={\nu \choose e}_L \qquad , \qquad \Psi_R={N_e \choose e}_R} 
\end{equation} 

\begin{equation} 
{q_L={u \choose d}_L \qquad , \qquad q_R={u \choose d}_R} 
\end{equation} 
and the quantum numbers are given in Table (5).

\begin{table}[!t]\label{panqueca}
\begin{center}
\begin{tabular}{||c||c||c||c||c||}
\hline States &  $I_{3L}$ & $I_{3R}$  & $(B-L)/2$ & $Q$ \\ 
\hline $\nu_L$ &  $1/2$ & $0$ & $1/2$ & $0$ \\ 
\hline  $e_L$  & $-1/2$ & $0$ & $1/2$ & $-1$ \\ 
\hline  $N_{eR}$ &  $0$ & $1/2$ & $1/2$ & $0$ \\ 
\hline  $e_R$ &    $0$ & $-1/2$ &  $1/2$ & $-1$ \\ 
\hline  $u_L$ &  $1/2$ & $0$ & $1/6$ & $2/3$ \\ 
\hline  $d_L$ &    $-1/2$ & $0$ &  $1/6$ & $-1/3$ \\ 
\hline  $u_R$ &  $0$ & $1/2$ & $1/6$ & $2/3$ \\ 
\hline  $d_R$ &    $0$ & $-1/2$ &  $1/6$ & $-1/3$ \\ 
\hline 
\end{tabular} 
\end{center}
\caption{Quantum numbers for left and right states in symmetric left-right model.}
\end{table}

\smallskip
\begin{table}[!t]\label{petit gateau}
\begin{center}
\begin{tabular}{||c||c||c||}
\hline  Couplings &  $g_V$ & $g_A$ \\ 
\hline  $Z\nu\nu$ & $1$ & $-1$ \\ 
\hline  $Zee$ & $-1+4sin\theta_W$ & $1$ \\ 
\hline  $Zuu$ & $3-8sin\theta_W$  & $-3$ \\ 
\hline  $Zdd$ & $-3+4sin\theta_W$ & $3$  \\ \hline
\hline  Couplings &  $g'_V$ & $g'_A$ \\ 
\hline  $Z'\nu\nu$ & $1$ & $-1$  \\ 
\hline  $Z'ee$ & $3$ &  $1$\\ 
\hline  $Z'uu$ & $-5$ & $-3$ \\ 
\hline  $Z'dd$ & $1$ & $3$ \\ 
\hline 
\end{tabular} 
\end{center}
\caption {Couplings between the neutral gauge bosons $Z$ and $Z'$ and the ordinary fermions in symmetric left-right model (first family).}
\end{table}

\par

\smallskip

\par
We can rewrite the gauge couplings in terms of a mixing angle as 
\begin{equation}
g= \frac{e} {\sin \theta_W}
\end{equation}

and

\begin{equation}
g'= \frac{e}{\sqrt{\cos 2\theta_W}}. 
\end{equation}

The neutral current Lagrangian that describes the interactions between the ordinary matter with $Z$ and $Z'$ boson contributions is

\begin{eqnarray}
{\cal L_{NC}} & = & \frac{e}{4 \sin \theta_W \cos\theta_W}\bar \Psi_i \gamma^{\mu}\lbrace 
(1- \gamma_5)I_{3_{L}}- Q\sin ^2 \theta_W\rbrace\Psi_i Z_{\mu}\nonumber\\
 \nonumber \\
& + & \frac{e}{\sin \theta_W \cos\theta_W} \frac{1}{\sqrt{\cos 2\theta_W}}  \bar \Psi_i \gamma^{\mu} \lbrace \sin ^2\theta_W(I_{3L}\frac{(1- \gamma_5)}{2}- Q \sin ^2\theta_W) 
\nonumber\\ \nonumber\\
& + & \cos^2\theta_W(I_{3R}\frac{(1 + \gamma_5)}{2}- Q \sin ^2 \theta_W)\rbrace\Psi_i Z'_{\mu} 
\end{eqnarray} 
and the resulting couplings are shown in Table (6).

\smallskip 

\begin{table}[!t]\label{caramelo}
\begin{center}
\begin{tabular}{||c||c||c||c||}
\hline \hline
\multicolumn{4}{|c|}{Couplings}  \\ \hline \hline
Models &  $g'_V$ & $g'_A$ & $g'_V/g'_A$\\ 
&      &     &   \\ \hline
\hline  SLRM   &    -0.08 & -0.54 & 0.15 \\
&      &     &   \\ \hline
\hline  MLRM      &  3    & 1 & 3\\
&      &     &   \\ \hline
\end{tabular} 
\end{center}
\caption{Couplings $g'_V$ and $g'_A$ of a new $Z'$ in mirror left-right model (MLRM) and symmetric left-right model (SLRM) and the ratio $g'_V/g'_A$ in both models. ($\sin^2\theta_W=0.23$).}
\end{table}

\smallskip

\begin{table}[!t]\label{manga}
\begin{center}
\begin{tabular}{||c|c|c|c||}
\hline \hline
\multicolumn{4}{|c|}{Couplings}  \\ \hline \hline
 &  $Z'^2$ & $Z'Z$ & $Z'A$\\ 
&      &     &   \\ \hline
$\chi^{02}_R$   &  $(gR_{22}+g'R_{32})^2$  & 
$2g^2 R_{21}R_{22} + 2g'^2 R_{31}R_{32}$ & $2g^2R_{22}R_{32} + 2g'^2R_{32}R_{33}$ \\
&      &  $-2g'g(R_{22}R_{31}+R_{32}R_{21})$   &  $-2gg'(R_{22}R_{33}+R_{32}R_{33})$  \\ \hline
\hline  $\chi^{02}_L$   &  $0$  & $-2g g'R_{11}R_{32} $ & $-2g g'R_{13}R_{32} $ \\
&      &     &   \\ \hline
\end{tabular} 
\end{center}
\caption{Couplings between scalar and gauge bosons.}
\end{table}

\smallskip

\par
In Table (7) we show the most important difference between the two models: the coupling of a new 
$Z'$ and ordinary charged leptons. The MLRM coupling is dominantly axial, whereas the SLRM is dominantly vectorial. This property will give different asymmetries, as will be shown in the next section.
\par

The Particle Data Group, in its 2002 edition \cite{RPP}, summarizes the
present data from low energy lepton interaction, lepton-hadron collisions
and the high precision data from LEP and SLAC. They also present the
experimental averages for the $g_V$ and $g_A$ couplings for charged and
neutral leptons. The most stringent bounds  come from the effective
coupling of the Z to the electron neutrino, $g_{exp}^{\nu e} =0.528 \pm
0.085$ and $\Gamma^{inv}_{exp}(Z)=499.0 \pm 1.5$ MeV, to be compared
with the standard model predictions  $g_{SM}=0.5042$ and
$\Gamma^{inv}_{SM}(Z)=501.65 \pm 0.15$ MeV. For the
muon neutrino coupling with the $Z$ boson, the Particle Data Group  quotes $g_{exp}^{\nu \mu} =0.502 \pm 0.017$. 
We have performed a fit to these  data, using the standard model predictions, and we find that deviations from the standard model must be bounded, at $95\%$ confidence level by: 

\bigskip

\begin{equation}
(\omega^2 \sin^4 \beta ) < 10^{-4}.
\end{equation}

\par

This bound is consistent with the present experimental constraint on the $\rho$ parameter.
With the value for $\sin\beta$ given in equation (3.3), we have the bound

\begin{equation}
v_R > 30 \, v_L.
\end{equation}

\par 
For the new $Z'$ mass we have

\begin{equation}
M_{Z'} > 800 \quad GeV
\end{equation} 
and the $Z'$ mass is the only new unknown parameter.

This value is a little above the present experimental bounds on new gauge bosons searches done by the CDF and DZero collaborations \cite{FAB} at Fermilab.\par

The $Z'$ total width in MLRM is $\Gamma_{Z'} \simeq  6.80\times 10^{-3} M_{Z'}$ and
$\Gamma_{Z'} \simeq  2.15\times 10^{-2} M_{Z'}$ in SLRM, 3 times larger than the previous model. The new $Z'$ decays can have contributions from many channels. For the channels $Z' \longrightarrow f \bar f$ with $"f"$ any of the presently known fermions we can compute all the decay ratios using Tables (4) and (6). A second group of decay channels comes from the triple and quartic vertices from Tables (1) and (2). All these channels give small contributions relative to the fermionic channels. The same
suppression is present in the scalar and neutral gauge bosons couplings as shown in Table (8). For example, the decay $Z' \longrightarrow  Z +  \chi_L +  \chi_L$ with $M_{\chi_L}= 150$ GeV  has a partial width $\Gamma_{Z'}= 2.46 \times 10^{-3}$ GeV for $M_{Z'}= 800$ GeV and $\Gamma_{Z'}= 2.67\times 10^{-1}$ GeV for $M_{Z'}=3$ TeV. In mirror models we can have new heavy fermions coupled to the new neutral current. These new exotic channels can have important phase suppression factors depending on their masses. Since these contributions depend on unknown parameters, we will not take them into account.
 
{\section {Results}}

\begin{figure}[t]
\resizebox{1.0\hsize}{!}{\includegraphics*{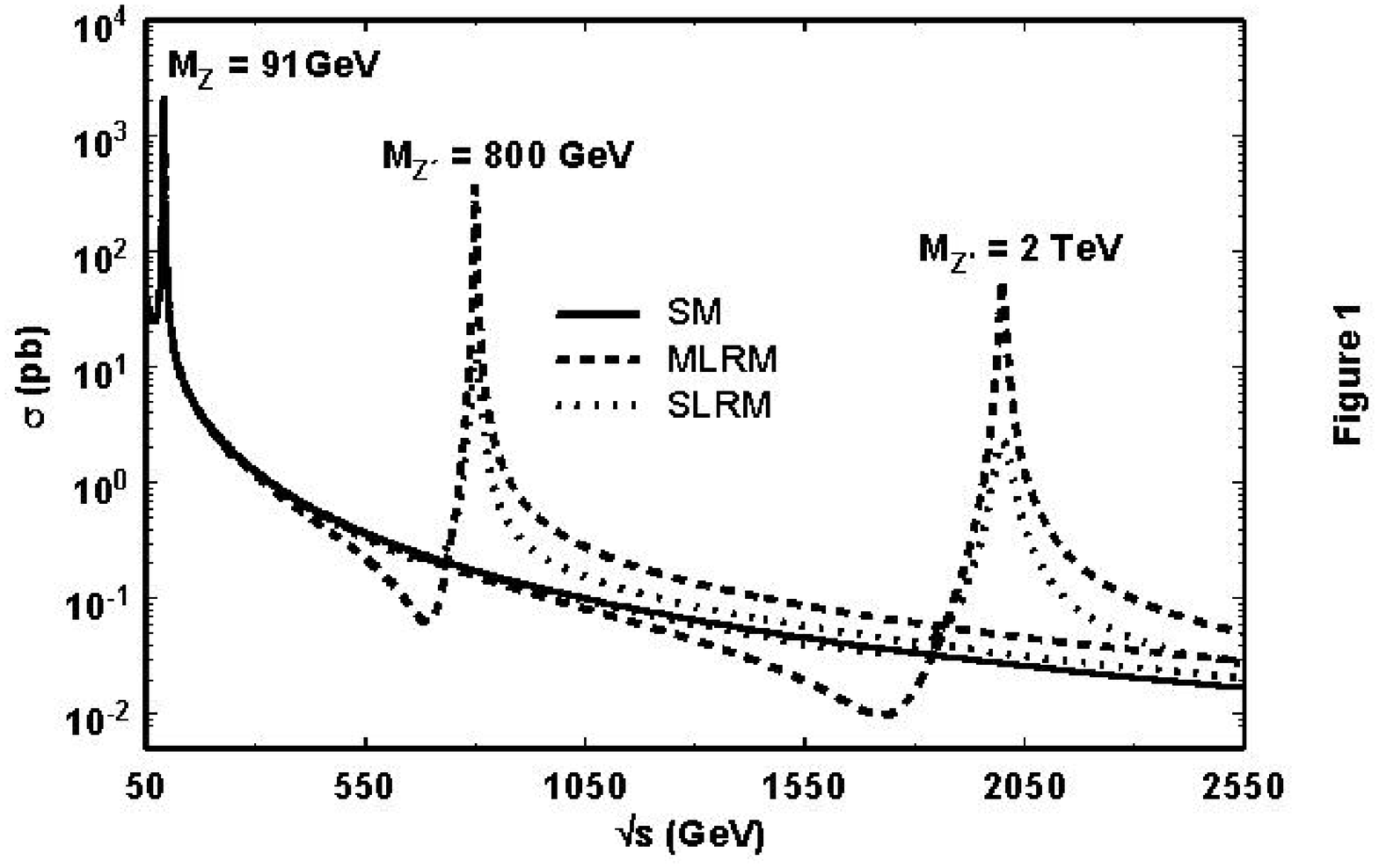}}
\caption{Total cross sections for muon pair production ${e^+}{e^-} \longrightarrow \mu^+ \mu^-$ versus ${\sqrt s}$ for standard model (SM), mirror left-right model (MLRM) and symmetric left-right model (SLRM).}
\end{figure}

\begin{figure}[t]
\resizebox{1.0\hsize}{!}{\includegraphics*{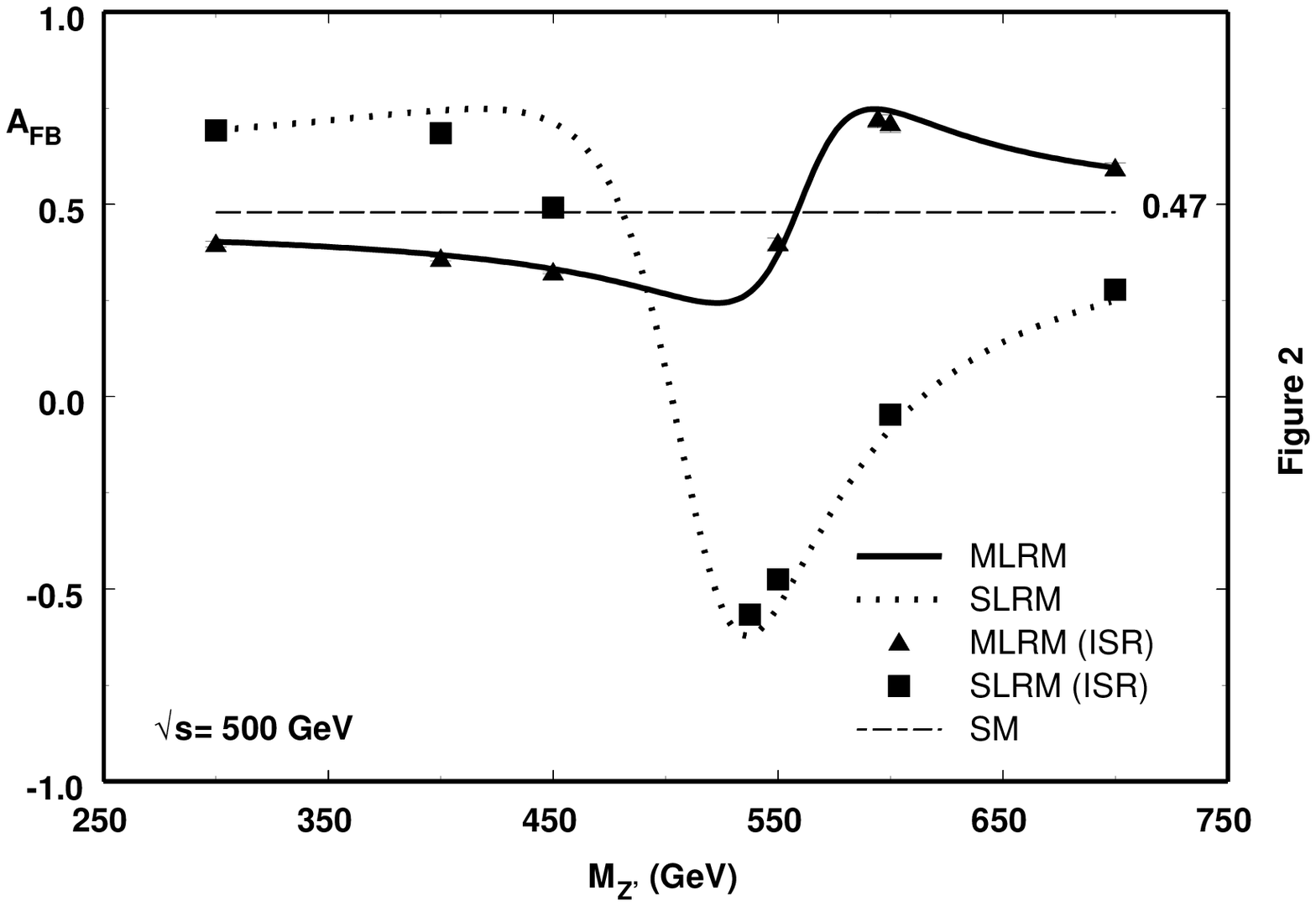}}
\caption{The forward-backward asymmetry in the process ${e^+}{e^-} \longrightarrow \mu^+ \mu^-$ 
for SM, SLRM and MLRM versus $M_{Z'}$ for TESLA ($\sqrt s = 500$ GeV).}
\end{figure}

In this section we present the total cross sections, angular distributions
and asymmetries for muon pair production in $e^+e^-$ annihilation,
comparing the signals from MLRM and SLRM with the standard model (SM) background. A Monte
Carlo program was written to generate events at a fixed c.m. energy 
$\sqrt{s}$. To be more specific, three energy values are considered in this
paper, $500$ GeV, $1$ TeV and $3$ TeV, which are appropriate for the TESLA at DESY, 
NLC at SLAC \cite{NLC} and CLIC at CERN \cite{BLA} respectively. In these
high-energy colliders, the incoming electrons and positrons radiate
photons, giving rise to the so-called initial state radiation (ISR), which
leads to an effective energy of the annihilation process smaller than the 
nominal c.m. energy of the colliding beams. In order to correct for ISR, 
the actual cross sections are written as convolutions of the Born cross 
sections for muon-pair production, with structure functions for the incoming
electron and positron beams. For these structure functions we follow the 
prescription of reference \cite{SKR}. The simulated events were selected by a
cut $\theta_{acol}< 10^{\circ}$ on the acollinearity angle of the
final-state muons, which are no longer produced back-to-back on account of
ISR. Both muons were also required to be detected within the polar angle
range $\vert\cos{\theta}\vert < 0.995$, where $\theta$ is the angle of either of the
muons with respect to the direction of the electron beam. For the
numerical calculations, we used $M_Z=91.1874$ GeV, $\Gamma_Z=2.496$ GeV,
$\alpha(M_Z^2) = 1/128.5$ and $\sin^2{\theta}_W = 0.23105$. Fermion masses
were set to zero. All the calculations involving unpolarized beams were cross-checked 
with CompHEP \cite{HEP}.\par

\begin{figure}[t]
\resizebox{1.0\hsize}{!}{\includegraphics*{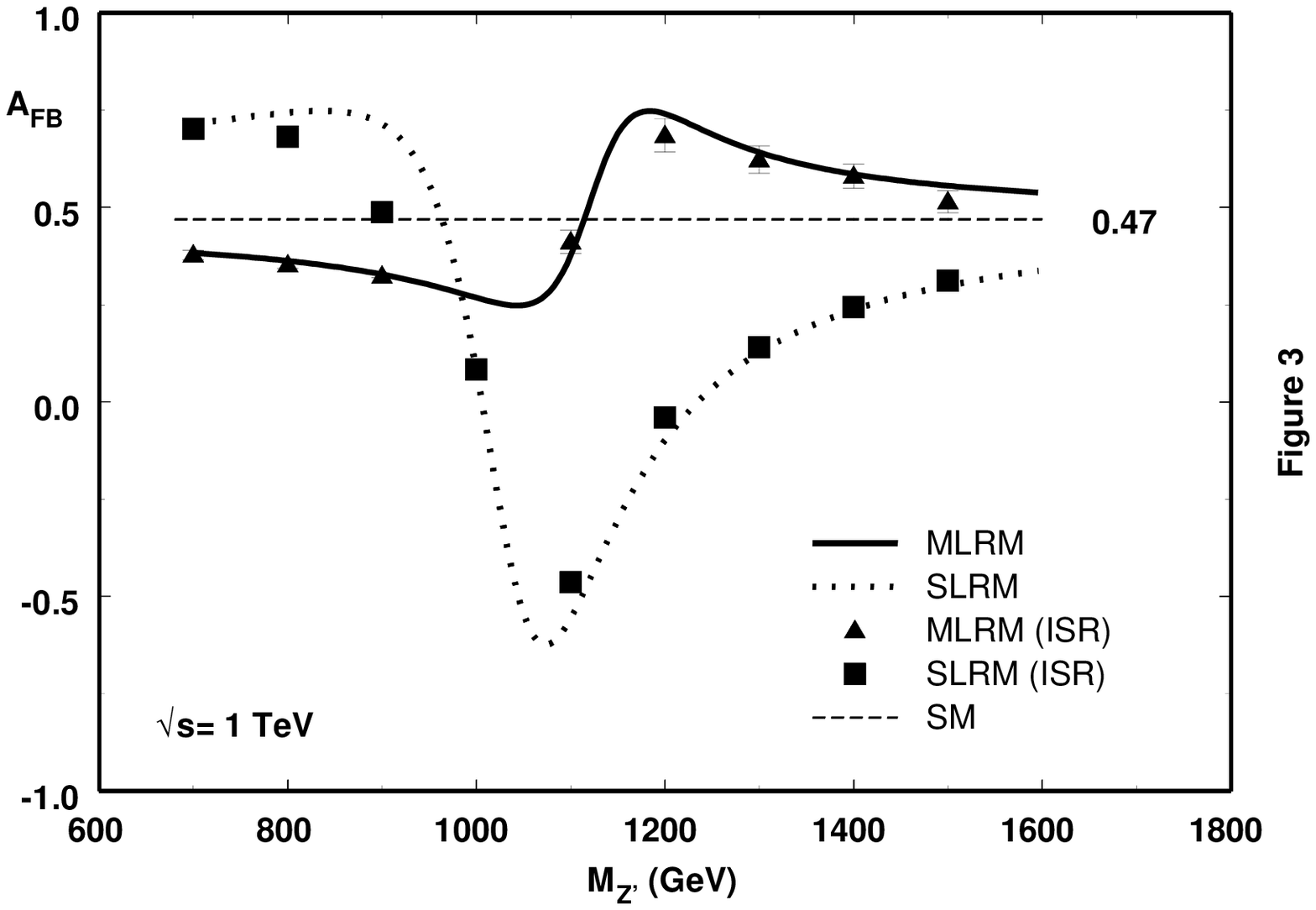}}
\caption{The forward-backward asymmetry in the process ${e^+}{e^-} \longrightarrow \mu^+ \mu^-$ 
for SM, SLRM and MLRM versus $M_{Z'}$ for NLC ($\sqrt s = 1$ TeV).}
\end{figure}

\begin{figure}[t]
\resizebox{1.0\hsize}{!}{\includegraphics*{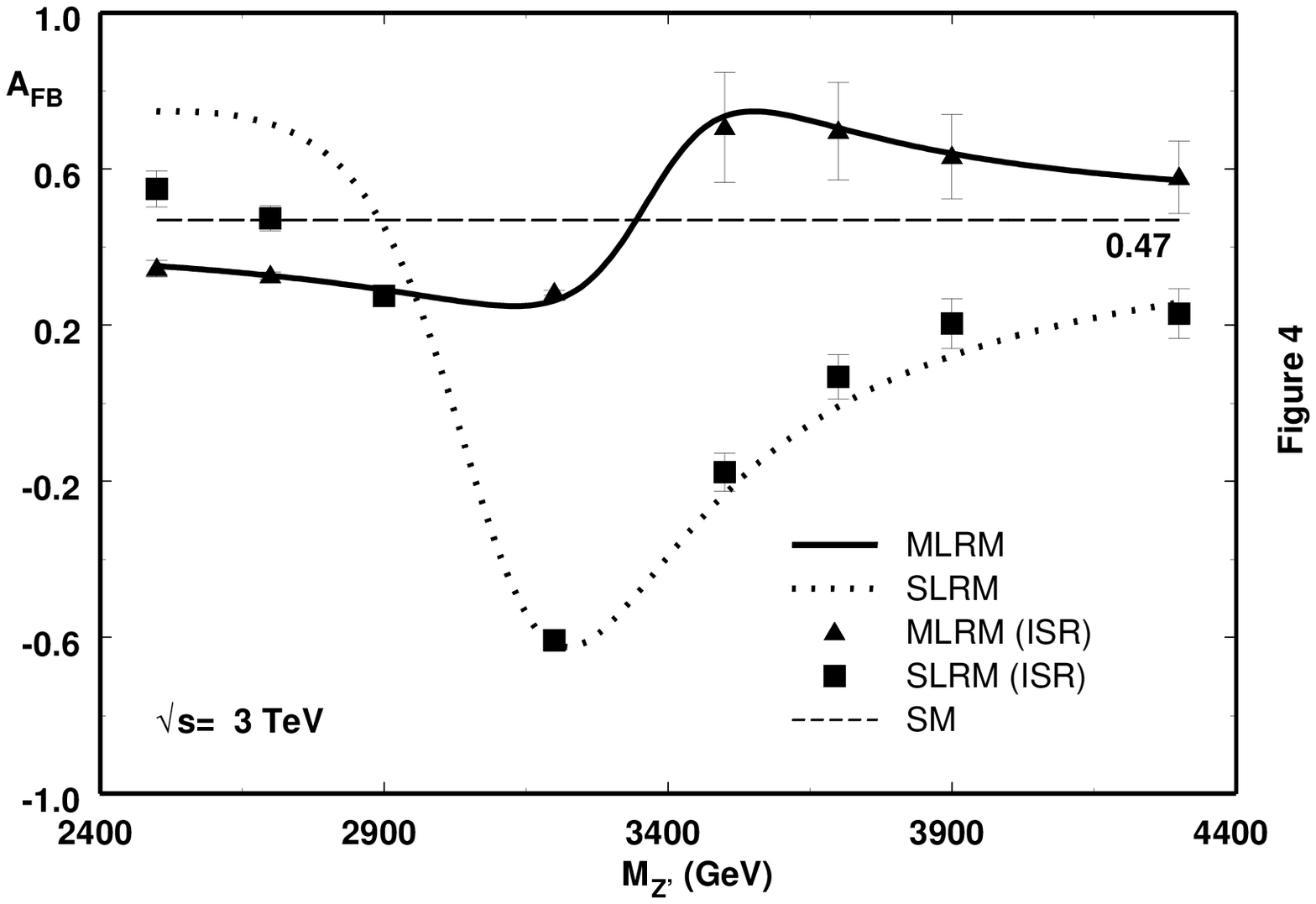}}
\caption{
The forward-backward asymmetry in the process ${e^+}{e^-} \longrightarrow \mu^+ \mu^-$ 
for SM, SLRM and MLRM versus $M_{Z'}$ for CLIC ($\sqrt s = 3$ TeV).
}
\end{figure}

In Figure 1 we show the total cross section without ISR for the process $e^+ + e^- 
\longrightarrow \mu^+ + \mu^- $, as a function of the c.m. energy, for SLRM
and MLRM. The SM cross section is also shown for comparison. Two
different values of $M_{Z^\prime}$ are considered, namely $M_{Z^\prime}= 800$
GeV and $M_{Z^\prime} = 2$ TeV. The expected resonance peaks associated
with these $M_{Z^\prime}$ values are clearly shown in the picture, as well
as the $Z^0$ SM peak. It is interesting to note that the peaks of the MLRM
cross sections are greater than those of the SLRM cross sections, because
the $Z^\prime$ total width is smaller in the MLRM. This property can be used to distinguish the two models. The presence of the new
neutral boson $Z^\prime$ is essential to preserve tree-level unitarity in
both extended models, leading to cross sections that fall to zero for
asymptotically high energies.\par
 
\begin{figure}[t]
\resizebox{1.0\hsize}{!}{\includegraphics*{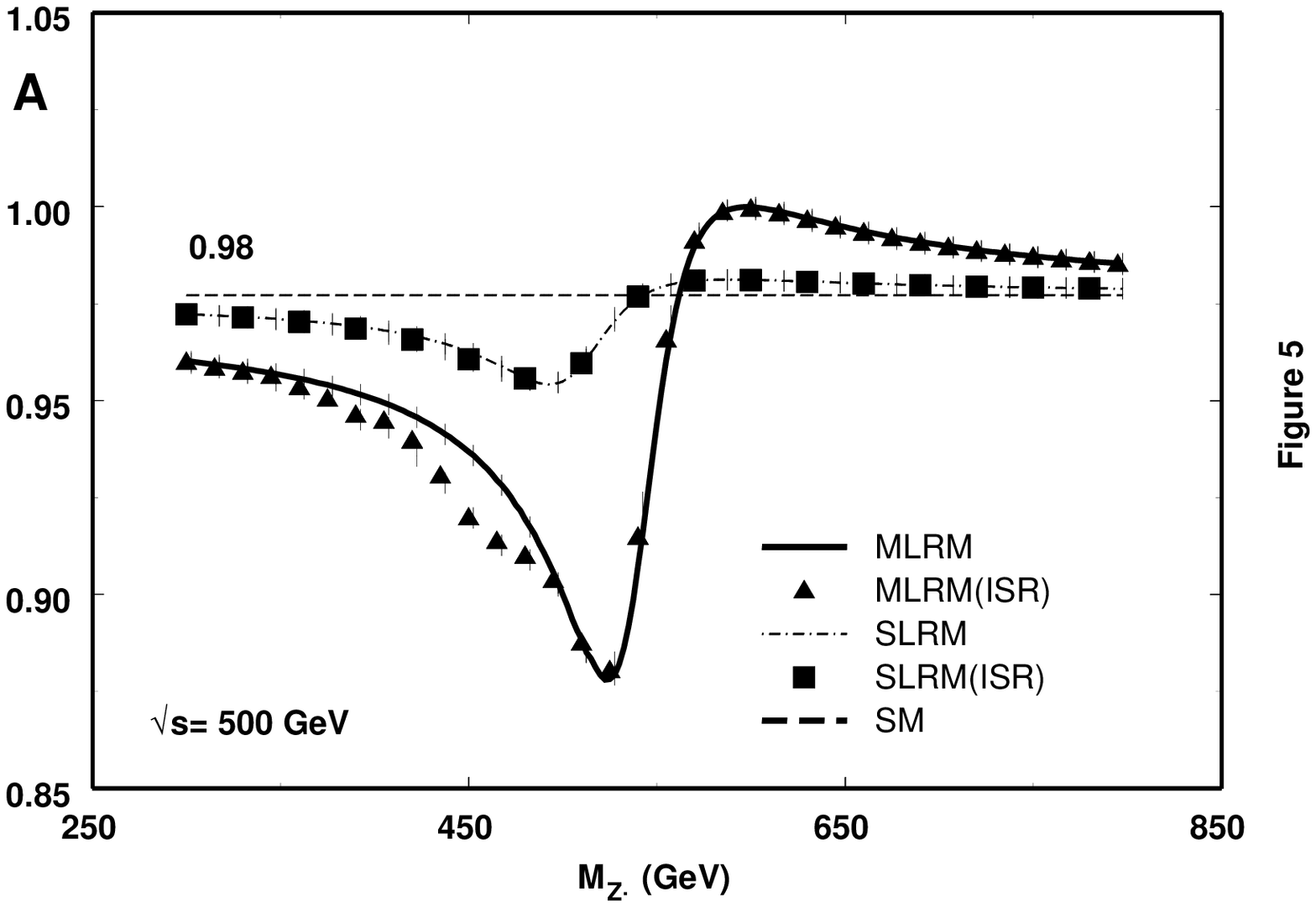}}
\caption{ The asymmetry ${\cal A}(P_-, P_+)$ in the process ${e^+}{e^-} \longrightarrow \mu^+ \mu^-$ for SM, SLRM and MLRM versus $M_{Z'}$ for TESLA ($\sqrt s = 500$ GeV). The longitudinal polarization of the electron and positron beams were taken to be $-90\%$ and $60\%$ respectively.}
\end{figure}

\begin{figure}[t]
\resizebox{1.0\hsize}{!}{\includegraphics*{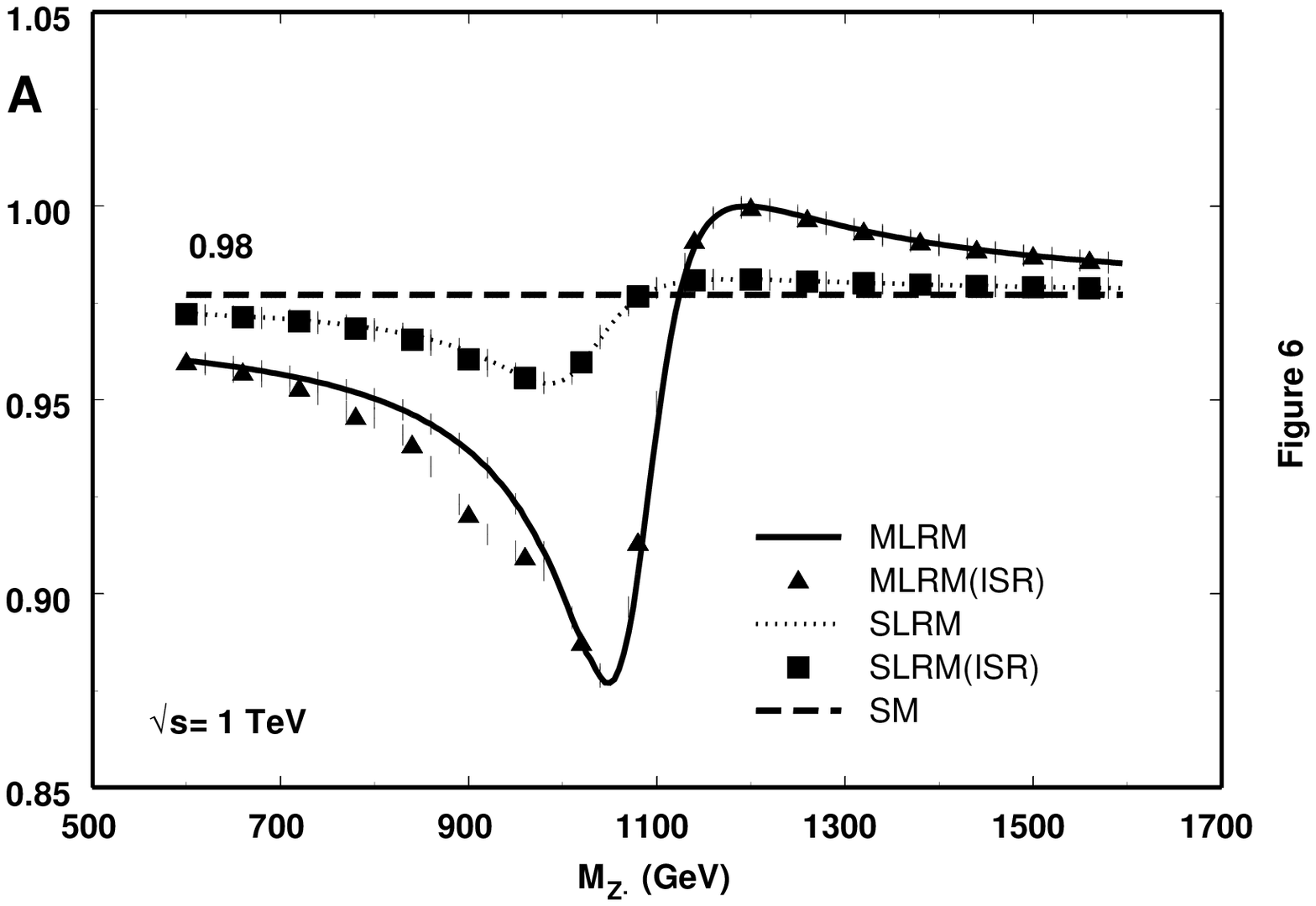}}
\caption{
The asymmetry ${\cal A}(P_-, P_+)$ in the process ${e^+}{e^-} \longrightarrow \mu^+ \mu^-$ for SM, SLRM and MLRM versus $M_{Z'}$ for NLC ($\sqrt s = 1$ TeV). The longitudinal polarization of the electron and positron beams were taken to be $-90\%$ and $60\%$ respectively.
 }
\end{figure}

Next we look at the dependence of the forward-backward asymmetry
$A_{FB}$ on $M_{Z^\prime}$. Figures 2, 3 and 4 show the corresponding curves
for the collider energies $500$ GeV, $1$ TeV and $3$ TeV respectively, and the points 
indicate how the ISR affects the asymmetry. In each case the error bars
represent the statistical errors for an integrated luminosity of $500$ fb$^{-1}$. The 
forward-backward asymmetry is quite sensitive to
$M_{Z^\prime}$, and can also be used to distinguish MLRM from SLRM.\par

\begin{figure}[t]
\resizebox{1.0\hsize}{!}{\includegraphics*{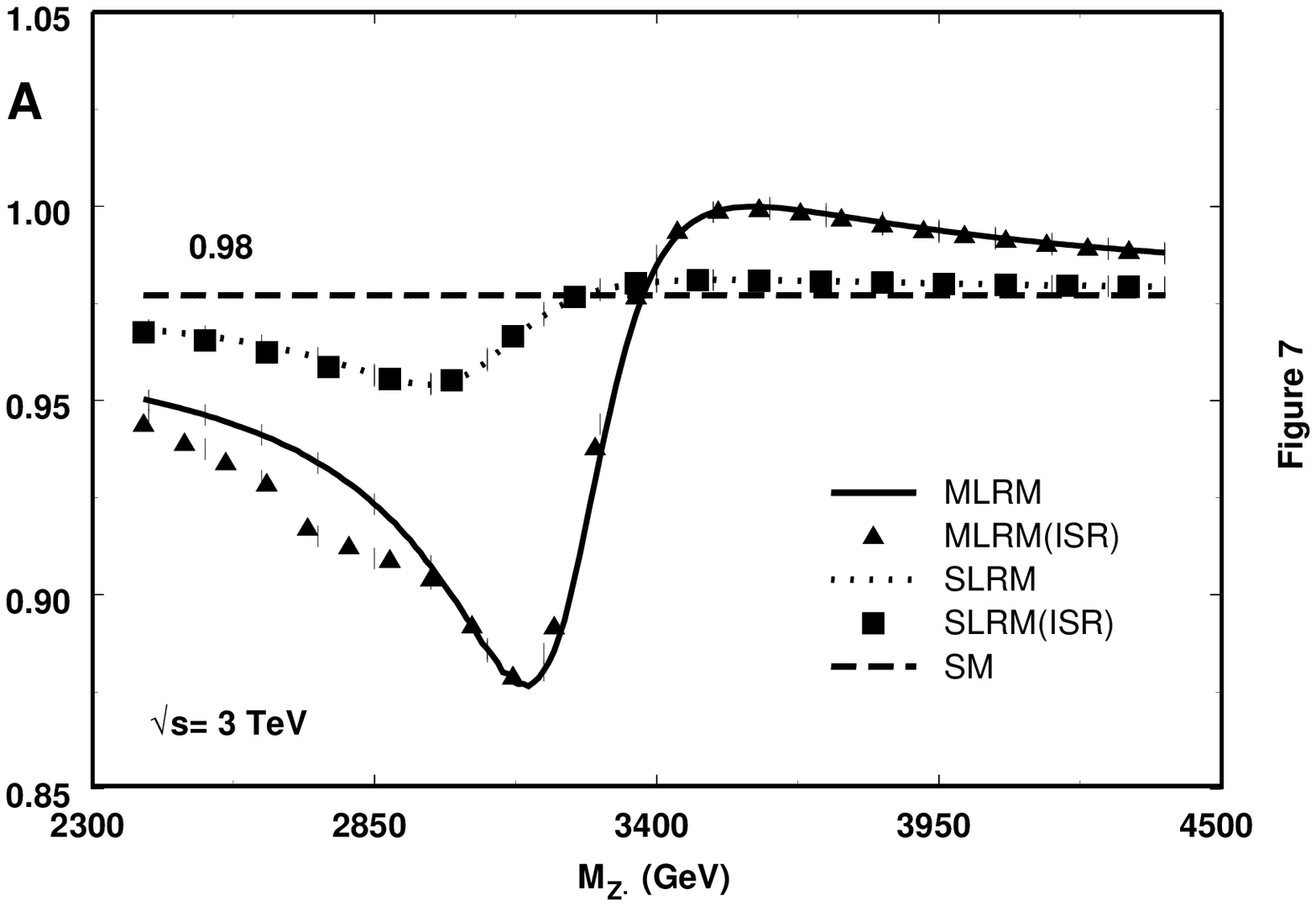}}
\caption{
The asymmetry ${\cal A}(P_-, P_+)$ in the process ${e^+}{e^-} \longrightarrow \mu^+ \mu^-$ for SM, SLRM and MLRM versus $M_{Z'}$ for CLIC ($\sqrt s = 3$ TeV). The longitudinal polarization of the electron and positron beams were taken to be $-90\%$ and $60\%$ respectively.
 }
\end{figure}

\begin{figure}[t]
\resizebox{1.0\hsize}{!}{\includegraphics*{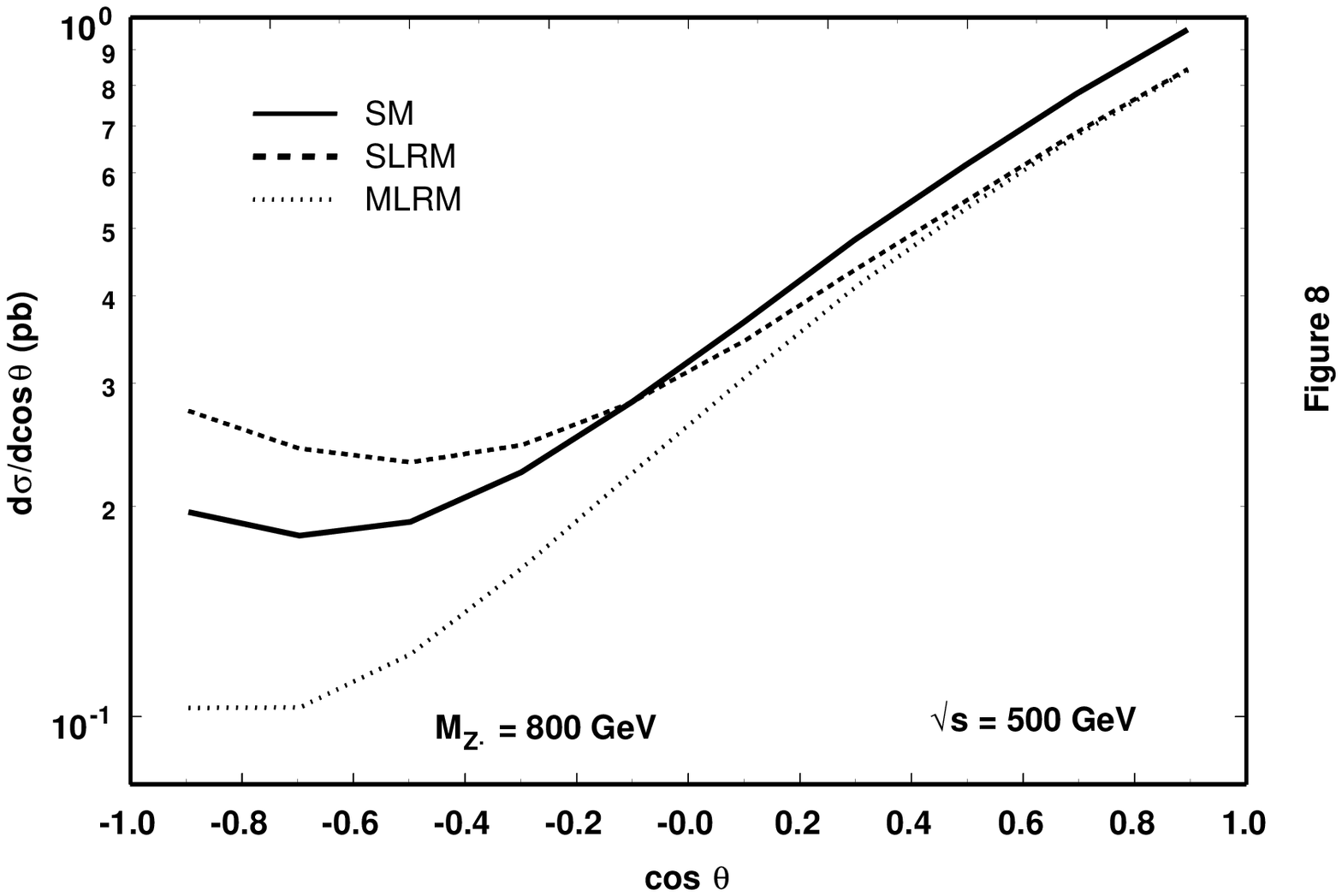}}
\caption{ Angular distributions of the $\mu^-$ in the process ${e^+}{e^-} \longrightarrow \mu^+ \mu^-$ for SM, SLRM and MLRM considering $M_{Z'}= 800$ GeV, $P_{-} = -90\%$ and $P_{+} = 60\%$ for TESLA ($\sqrt s = 500$ GeV).}
\end{figure}

Beam polarization is expected to play a very important role at the
future linear collider facilities. With longitudinally polarized electron
and positron beams one can effectively  enhance the signals of interest,
and suppress inconvenient backgrounds, and thus increase the sensitivity
of spin-dependent observables to deviations from the SM
predictions. Experts usually believe that it should not be too difficult
to produce electron beams whose degrees of polarization exceed $90\%$. As
a matter of fact, electron beam polarization routinely reaches values
around $80\%$ at SLAC. Several schemes have been devised to produce
polarized positron beams in a linear collider. Although these techniques
remain untested, simulations suggest that it is feasible to reach a degree
of positron polarization of $60\%$. In all the calculations considered in
the following, the degrees of polarization of the electron and positron
beams were taken to be $P_{-}= -90\%$ and $P_{+} = 60\%$ respectively. To
illustrate the importance of beam polarization, it suffices to say that
for $\sqrt{s} = 500$ GeV, the polarized cross section $\sigma(P_{-},P_{+})
= \sigma(-0.9,0.6)$ is essentially double the unpolarized cross section
$\sigma(0,0)$ . Here we define an asymmetry ${\cal{A}}(P_{-},P_{+})$, in
terms of the degrees of polarization $P_{\pm}$ of the electron and
positron beams, and the helicity cross sections:\par

\begin{figure}[t]
\resizebox{1.0\hsize}{!}{\includegraphics*{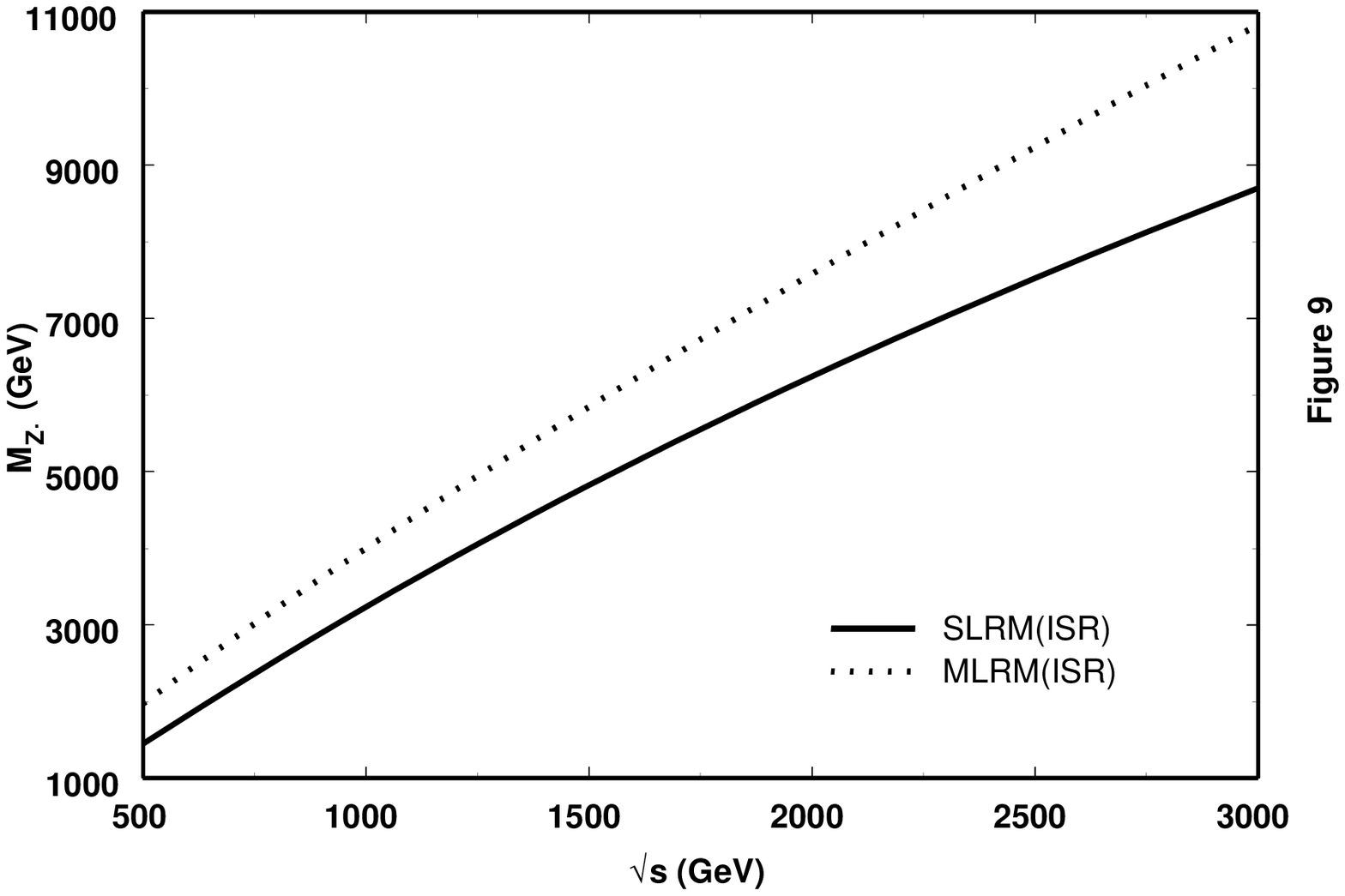}}
\caption{
$Z'$ mass bound with $95\%$ C.L. as a function of $\sqrt s$ for MLRM and SLRM. 
 }
\end{figure}

\smallskip
\begin{equation}
{\cal{A}}(P_{-},P_{+}) = \displaystyle \frac {(1-P_{-})(1+P_{+})\sigma_{-+} - (1+P_{-})(1-P_{+})
\sigma_{+-}}  {(1-P_{-})(1+P_{+})\sigma_{-+} + (1+P_{-})(1-P_{+})\sigma_{+-}} \quad ,
\end{equation} 
where the first (second) subscript in $\sigma_{\pm \mp}$ refers to the
electron (positron) helicity. The parity violating  left-right
asymmetry $A_{LR} = (\sigma_L - \sigma_R)/(\sigma_L + \sigma_R)$ can
be easily obtained from $\cal{A}(P_{-},P_{+})$ through the relation

\smallskip
\begin{equation}
A_{LR} = \displaystyle \frac {{\cal{A}}(P_{-},P_{+})-P_{eff}}
{1 - P_{eff} \cdot {\cal{A}}(P_{-},P_{+})} \quad ,     
\end{equation}
with the effective polarization defined as $P_{eff}= (P_{+}-P_{-})/
(1 - P_{-}P_{+})$.

Figures 5, 6 and 7 display the behavior of $\cal{A}(-0.9,0.6)$ as a function
of $M_{Z^\prime}$, for the three energies under study. The differences
between MLRM and SLRM asymmetries are considerable, the more so in the 
resonance region, the deviations from the SM value being larger
for the MLRM over the whole $M_{Z^\prime}$ range. As expected, the
asymmetries in both models tend to the standard model value for 
$M_{Z^\prime} \gg \sqrt{s}$. If we take into account the lower bound in 
Equation (3.14), $M_{Z^\prime} > 800$ GeV, it seems unlikely that 
$\cal{A}(P_{-},P_{+})$ can be used as a measure of the deviations of these 
models from the SM at the first stage of TESLA, in which 
$\sqrt{s} \le 500$ GeV. For a possible TESLA extension, where the c.m. 
energy could reach $800$ GeV, detection of these deviations in 
${\cal{A}}(P_{-},P_{+})$ can not be excluded if $M_{Z^\prime}$ is close to 
the lower bound. For higher values of $\sqrt{s}$, as those of NLC (stage 2)
and CLIC, ${\cal{A}}(P_{-},P_{+})$ is sensitive to larger values of the 
$Z^\prime$ boson mass, as long as we exclude the asymptotic region 
$M_{Z^\prime} \gg \sqrt{s}$.
\par

In order to determine the discovery limits for a $Z^\prime$ boson via 
muon pair production, we compared the angular distribution
$d\sigma/d(cos\theta)$ predicted by each of the left-right models with the
corresponding SM expectation. Plots of the angular
distribution are shown in Figure 8 for the extended models and SM,
considering $M_{z^\prime} = 800$ GeV,
$P_{-} = -90\%$ and $P_{+} = 60\%$ for $\sqrt{s} = 500$ GeV. 
Assuming that the experimental data in
muon pair production will be well described by the standard model
predictions, we defined a one-parameter $\chi^2$ estimator
\smallskip
\begin{equation}
\chi^2(\xi) = \sum_{i=1}^{n_b} {\biggl( {N_i^{SM}- N_i^{LR} \over 
\Delta N_i^{SM}}\biggr)^2},
\end{equation}
\smallskip
where $N_i^{SM}$ is the number of SM events collected in the 
$i^{th}$ bin, $N_i^{LR}$ is the number of events in the $i^{th}$ bin as
predicted by the extended model, and $\Delta N_i^{SM} = 
\sqrt{(\sqrt {N_i^{SM}})^2 + (N_i^{SM}\epsilon)^2}$ the corresponding
total error, which combines in quadrature the Poisson-distributed statistical
error with the systematic error. We took $\epsilon = 5\%$ to correct for
those sources of systematic error not explicitly accounted for in our
calculation, such as the luminosity uncertainty, beam energy spread and
the uncertainty in the muon detection efficiency. The angular
range $\vert cos\,\theta \vert < 0.995$ was divided into $n_b= 10$ equal-width
bins, and the free parameter $\xi = 1/M_{Z^\prime}$ was varied
to determine the $\chi^2(\xi)$ distribution. The $95\%$
confidence level bound corresponds to an increase of the $\chi^2$ by
$3.84$ with respect to the minimum $\chi^2_{min}$ of the distribution.
Figure 9  represents the $95\%$ confidence limits on the
$(\sqrt{s},M_{Z^\prime})$ plane for TESLA, NLC and CLIC.\par

\bigskip
{\section {Conclusions}}

\par

We have presented an analysis of the effects of a new neutral gauge boson
$Z^\prime$ in muon pair production, at the next generation of linear
colliders, in the context of two extended electroweak models, the mirror left-right
model (MLRM) and the symmetric left-right model (SLRM). A number of observables that are
sensitive to the presence of such a gauge boson were studied in detail.
These observables were found to be useful to distinguish the two models,
should new physics associated with the $Z^\prime$ turn up at high mass
scales. Our simulations indicate that longitudinally polarized electron
and positron beams can significantly increase event rates and the
sensitivity of these observables to the presence of a new neutral gauge
boson. Starting from the angular distributions of the final $\mu^-$,  $95\%$ C. L. discovery limits on the $Z^\prime$ mass were derived for the new linear colliders, in terms of the available c.m. energies.
\bigskip

{\it Acknowledgments:} This work was partially supported by the
following Brazilian agencies: CNPq and FAPERJ.

\end{document}